\documentclass[twocolumn, amsmath,amssymb,aps,letter,floatfix]{revtex4-1}

\usepackage{float}
\usepackage{graphicx}
\usepackage{dcolumn}
\usepackage{bm}
\usepackage{fdsymbol}

\usepackage[usenames,dvipsnames]{color}
\usepackage{subfigure}
\usepackage{longtable}
\usepackage{multirow}
\usepackage{comment}
\usepackage{manfnt}
\usepackage{soul}
\usepackage[normalem]{ulem}
\usepackage{wasysym}

\usepackage{bbding}

\begin{document}

\preprint{APS/123-QED}

\title{Temperature Dependence of the Viscoelastic Properties of a Natural Gastropod Mucus by Brillouin Light Scattering Spectroscopy}

\author{Dillon F. Hanlon}
\email{dfh031@mun.ca}
\author{Maynard J. Clouter}
\author{G. Todd Andrews}

\affiliation{Department of Physics and Physical Oceanography, Memorial University of Newfoundland and Labrador, St. John's, NL, Canada, A1B 3X7}

\date{\today}

\begin{abstract}
Brillouin spectroscopy was used to probe the viscoelastic properties of a natural gastropod mucus at GHz frequencies over the range -11 $^\circ$C $\leq T \leq$ 52 $^\circ$C. Anomalies in the temperature dependence of mucus longitudinal acoustic mode peak parameters and associated viscoelastic properties at $T = -2.5^\circ$C, together with the appearance of a peak due to ice at this temperature, suggest that the mucus undergoes a phase transition from a viscous liquid state to one in which liquid mucus and solid ice phases coexist. Failure of this transition to proceed to completion even at -11 $^\circ$C is attributed to glycoprotein-water interaction.  The temperature dependence of the viscoelastic properties and the phase behaviour suggest that water molecules bind to glycoprotein at a temperature above the onset of freezing and that the reduced ability of this bound water to take on a configuration that facilitates freezing is responsible for the observed freezing point depression and gradual nature of the liquid-solid transition.

\begin{description}

\item[PACS numbers]
May be entered using the \verb+\pacs{#1}+ command.

\end{description}
\end{abstract}

\pacs{Valid PACS appear here}

\maketitle

\section{Introduction}
Gastropod mucus is a fascinating polymer hydrogel consisting primarily of long chains of tangled high molecular weight glycoproteins in 91 wt\% to 98 wt\% water \cite{denn1980,denn1984,verd1987}.  These glycoproteins, despite being present at low weight percentages, are anticipated to play a dominant role in many of the physical properties of this natural complex fluid.  This influence is expected to be especially apparent in the mucus viscoelasticity because these biopolymers expand in water and cross-link to form a gel \cite{denn1980}.  In fact, stress-strain measurements with strain rates $\leq 100$ Hz reveal non-Newtonian behaviour with the mucus displaying characteristics of a soft elastic solid at low strains and yielding at higher strains and presenting as a viscous liquid \cite{denn1980}.  Analogous measurements of the high-frequency ({\it i.e.}, MHz to GHz) viscoelasticity have not been performed, but based on studies of other complex fluids, probing the upper end of this range would be particularly valuable because it would lead to the discovery of links between the viscoelasticity and the dynamics and structure of the mucus \cite{liao1973,gram1995,ye1993}.  Quantification of the temperature dependence of the mucus viscoelastic properties in this frequency range would also be of considerable value because it would reveal other glycoprotein-induced anomalies and permit the phase behaviour of the mucus to be mapped. Such data is important for comparison to molecular dynamics simulations results on protein-water dynamics to inform refinement of models \cite{meis2013}.  Knowledge of the temperature dependence of gastropod mucus viscoelasticity is also important from a technological standpoint.  For example, synthetic mucus mimics are being developed for possible application as non-toxic adhesives to close internal wounds due to the ability of mucus to adhere to wet surfaces \cite{li2017} and solutions of so-called antifreeze glycoproteins are being investigated as potential replacements for conventional cryprotectants due to their ability to provide the required high viscosity at low concentrations \cite{bouv2003,kwan2020}. 

In this study, Brillouin light scattering spectroscopy was used to characterize the temperature dependence of the GHz-frequency viscoelastic properties of a natural snail mucus over the range -11 $^\circ$C to 52 $^\circ$C.  We report values for hypersonic velocity and attenuation, compressibility, complex longitudinal modulus, and apparent viscosity, and document a large and previously undiscovered glycoprotein-induced freezing point depression associated with an incomplete phase transition from the liquid state to one in which liquid mucus and solid ice coexist. These results complement previously published data on the low-frequency viscoelasticity of gastropod mucus and provide new insights into the physics of aqueous glycoprotein solutions and the role played by glycoproteins in the phase behaviour of these systems. In a broader context, the present study contributes new knowledge on phonon dynamics and phase transitions in complex fluids, and on water-macromolecule interaction vital to understanding the intricate behaviour of biological systems.

\section{Brillouin Scattering in Liquids}
Brillouin spectroscopy is a technique used to probe thermal acoustic waves by inelastic scattering of light. For a $180^{\circ}$ backscattering geometry as used in the present work, application of energy and momentum conservation to the scattering process reveals that the hypersound velocity, $v$, and frequency shift of the incident light, $f_B$, are related by  
\begin{equation}
v = \frac{f_B\lambda_i}{2n},
\label{eq:brillouineqn}
\end{equation}
where $n$ is the refractive index of the target material at the incident light wavelength $\lambda_{i}$.

For liquids, the longitudinal kinematic viscosity or apparent viscosity $\eta = 4\eta_{s}/3 + \eta_{b}$, where $\eta_{s}$ and $\eta_{b}$ are the shear and bulk viscosity, respectively, is related to the full width at half maximum of the Brillouin peak due to the longitudinal acoustic mode via \cite{rouc1976},
\begin{equation}
    \Gamma_B = \frac{16\pi^2 n^2}{\rho \lambda_i^2} \left[\eta + \frac{\kappa}{C_p}(\gamma - 1) \right], 
    \label{eq:FWHM}
\end{equation}
where $\rho$ is the density, $\kappa$ is the thermal conductivity, and $\gamma = C_p/C_v$ is the ratio of specific heat at constant pressure to that at constant volume.  The second term in the brackets of this expression is usually neglected for simple liquids, thus leaving the expression for viscosity \cite{rouc1976},
 \begin{equation}
\eta = \frac{\rho \lambda_i^2}{16 \pi^2 n^2}\Gamma_B.
    \label{eq:viscosity}
\end{equation}
The linewidth $\Gamma_B$ is determined by the time that a thermal density fluctuation interacts with the incident light and is therefore a measure of its lifetime or, equivalently, its attenuation \cite{dil1982}.

The complex longitudinal modulus, often referred to simply as the complex modulus, is related to the frequency shift and linewdith via  \cite{bail2020},
 \begin{equation}
    M = M^{\prime} + M^{\prime\prime}i = \rho \left(\frac{f_B \lambda_i}{2n}\right)^2 + 2\pi\left(\frac{\rho\lambda_i^2}{16 \pi^2 n^2}\Gamma_B \right) f_Bi,
    \label{eq:longmod}
\end{equation}
where the quantities in parentheses are the hypersound velocity and apparent viscosity given by Eqs. \ref{eq:brillouineqn} and \ref{eq:viscosity}, respectively.  $M^{\prime}$ is the storage modulus (or bulk modulus) and is a measure of how much energy is stored elastically in the system. The loss modulus, $M^{\prime\prime}$, is a measure of how much energy is lost through heat in the system.    

The frequency-independent sound absorption coefficient can also be expressed in terms of $f_B$ and $\Gamma_B$ through \cite{rouc1976},
\begin{equation}
\frac{\alpha}{f^2} = \frac{\Gamma_B}{2vf_B^2}.
\label{eq:absorp}
\end{equation}
Measurements of Brillouin peak frequency shift $f_B$ and linewidth $\Gamma_B$ thereby permit direct determination of hypersound velocity, apparent viscosity, complex longitudinal modulus, and sound absorption coefficient via Eqs. \ref{eq:brillouineqn} through \ref{eq:absorp}.

\section{Experimental Details}

\subsection{Sample Preparation}
The mucus used in the present study was a transparent  viscous natural snail mucus. The mucus was kept in a sealed container at room temperature until a sample was required for use in experiments.  A syringe was used to extract $\sim$5 ml of mucus from the parent container.  One end of a thin plastic tube was then attached to the tip of the loaded syringe while the other end was placed in the bottom of a sample cell consisting of a $\sim15$ cm-long, 6 mm OD/4 mm ID glass tube sealed by glassblowing at one end. By applying gentle pressure to the syringe plunger, mucus was slowly forced from the syringe barrel, through the tube, and into the sample cell.  This gradual extrusion helped minimize air bubble formation in the mucus during loading.  To allow space for expansion upon cooling near and at the mucus freezing point, the cell was only filled to half capacity ($\sim1$ cm$^{3}$ of mucus).  Once loaded, the top of the sample cell was sealed with a Teflon cap held in place with epoxy.  After loading, the glass cell was placed in a sample holder in preparation for Brillouin scattering experiments. 

\subsection{Temperature-Controlled Sample Chamber}
\begin{figure*}[ht]
\includegraphics[scale = 0.44]{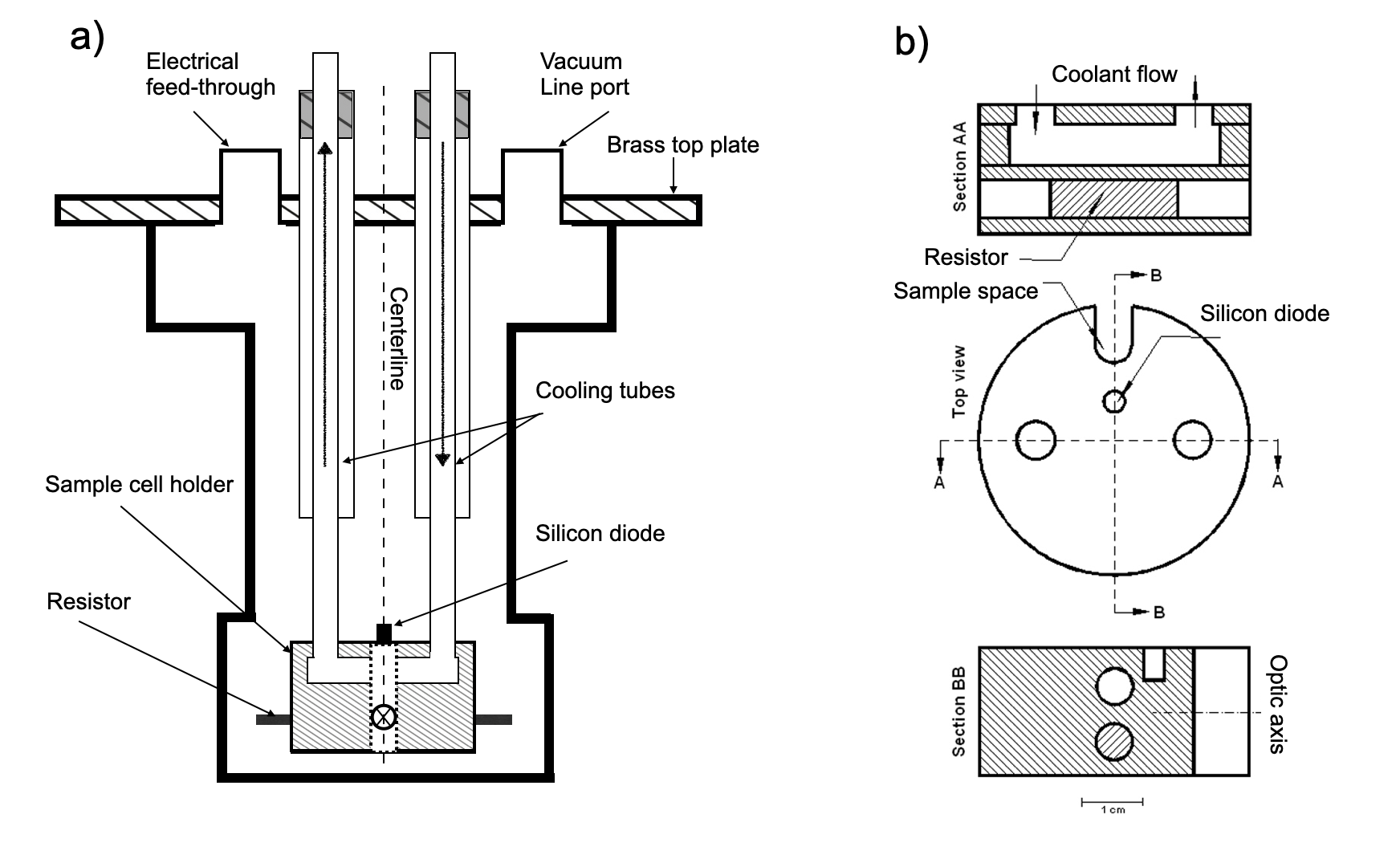}
  \caption{a) Schematic of the custom-built temperature-controlled sample chamber used in the present work with sample cell holder in place. b) Machine drawing of the sample cell holder shown in a).}
  \label{fig:TempCell}
\end{figure*}

Fig. \ref{fig:TempCell} shows schematic diagrams of the custom-built temperature-controlled sample chamber (Fig. \ref{fig:TempCell}a) and sample cell holder (Fig. \ref{fig:TempCell}b). The sample chamber was made primarily from aluminum and was designed to enclose the sample cell holder. The brass top plate had an electrical feed-through and a vacuum line port. It was important in this work that the sample chamber be evacuated to ensure the brass sample cell holder, and therefore the sample, was at the desired temperature with high precision. Brass was used for the sample cell holder due to its exceptional thermal properties. Cooling tubes in the sample chamber allowed for the circulation of antifreeze coolant through the sample holder which helped for cooling the holder and for temperature stabilization. A 50 $\Omega$ resistor was placed in a vacant space in the sample holder to heat up the holder and to also assist with stabilization. A silicon diode was used to obtain the temperature of the sample cell holder. A glass sample cell which the sample resides in was held in place in the sample space shown in the sample cell holder (vertical dashed line, Fig. \ref{fig:TempCell}a) and the sample space labelled in Fig \ref{fig:TempCell}b. Thermal paste was used to ensure adequate contact between the glass sample cell and the brass sample cell holder.

The temperature of the sample cell holder is controlled by use of a heater and cooling bath system. Using a  cryogenic temperature controller from Lakeshore Cryotronics. Depending on the desired temperature of the experiment, either the heater or the cooling bath is initially turned on.  Once the desired temperature is reached, the heater and refrigeration bath will work together to stabilize the sample cell holder. If the sample cell holder surpasses the desired temperature (voltage set-point) the heater will turn off and the cooling bath will turn on. Similarly, if the sample cell holder gets too cold, the bath will turn off and the heater will turn on.  A full description of the temperature control operation has been explained previously \cite{gagn1988, gamm1983_2}.

Accuracy of the temperature controlled sample holder was measured by collecting Brillouin spectra of water between  -5$^\circ$C $\leq T \leq$ 5$^\circ$C, with a focus on obtaining data near $T = 0$ $^\circ$C. The temperature at which spectral peaks due to ice appear was deemed the freezing point of water. The uncertainty in temperature reading from the diode was attributed to the difference between the freezing point of water (when ice appeared) and the freezing point of water found in the literature \cite{brownridge2011does}.  Brillouin peaks due to ice were observed in the spectra at a temperature of -0.4$^\circ$C, thus the uncertainty associated with the temperature reading in this work was $\pm$ 0.4$^\circ$C.

\subsection{Brillouin Light Scattering Apparatus}
Brillouin spectra were obtained using a $180^\circ$ backscattering geometry using the set-up shown in Ref. \cite{andr2018}. Incident light of wavelength $\lambda_{i} = 532$ nm and power of 100 mW was provided by a Nd:YVO$_4$ single mode laser. A high-quality anti-reflection-coated camera lens of focal length $f=5$ cm and $f/\# = 2.8$ served to both focus incident light onto the sample and to collect light scattered by it.  After exiting this lens, the scattered light was focused by a 40 cm lens onto the 450 $\mu$m-diameter input pinhole of an actively-stabilized 3+3 pass tandem Fabry-Perot interferometer (JRS Scientific Instruments) by which it was frequency-analyzed.  The free spectral range of the interferometer was set to 30 GHz and the finesse was $\sim100$. The light transmitted by the interferometer was incident on a 700 $\mu$m pinhole and detected by a low-dark count ($\lesssim 1$ s$^{-1}$)  photomultiplier tube where it was converted to an electrical signal and sent to a computer for storage and display. 

\begin{table*}[t] \footnotesize
\caption{Refractive index and density of snail mucus, water, and ice Ih.  Lone entries in the `Refractive Index' and `Density' columns indicate that the temperature dependence of the quantity was not known and was therefore taken to be constant and equal to the value shown over the temperature range probed in the present work.}
\begin{ruledtabular}
\begin{tabular}{ccc}
Substance & Refractive Index & Density \\ 
      & [@ 532 nm] & [kg/m$^3$] \\ \hline
 & & \\
Mucus & 1.34 \cite{gugl2021} & 1040 [Present Work] \\ [0.5cm] \hline
 & & \\
Water & $\displaystyle \left(\frac{n^2-1}{n^2+2}\right)\frac{1}{\rho} = 2.180454 \times 10^{-4} + 9.746345 \times 10^{-9}\rho$ & $T\geq 0$ $^\circ$C:  Values in Table I \cite{ocon1967} \\ [0.2cm]
  & \hspace*{3.1cm} $-1.286164 \times 10^{-8}T -1.666262 \times 10^{-11}\rho^2$ \cite{harv1998} &   $T < 0$ $^\circ$C: Values in Table II \cite{hare1987}  \\ [0.5cm] \hline
   & & \\
Ice Ih & 1.3117 at -7 $^\circ$C by interpolation of data in Table I \cite{hare1984} & $\rho = \rho_0/[1 + 1.576 \time 10^{-4}T - 2.778 \times 10^{-7}T^2$ \\ [0.2cm]
 & & \hspace*{1.3cm} $+ 8.850 \times 10^{-9}T^3 - 1.778 \times 10^{-10}T^4]$ \cite{gamm1983} \\
\end{tabular}
\end{ruledtabular}
\label{tab:n_rho}
\end{table*}

\subsection{Ancillary Quantities}
To determine viscoelastic properties from Brillouin data the refractive index and density of the mucus, water, and ice were required.  The density of the mucus was found to be 1040 $\pm$ 30 kg/m$^3$ by measuring the mass of a precisely known volume (100 ml) using a (Secura\textregistered, Satorius) microbalance and was taken to be constant over the temperature range probed in these experiments.  This value is in good agreement with published values \cite{gugl2021}.  The refractive index of mucus was taken to be $n=1.34$ \cite{gugl2021}.  For water, the refractive index and density were obtained from an empirical expression in Ref. \cite{harv1998} and tabulated values in Refs. \cite{ocon1967} and \cite{hare1987}, respectively.  For ice, the refractive index at 532 nm was determined by interpolation of data in Table I of Ref. \cite{hare1984} and the density using an empirical expression from Ref. \cite{gamm1983}. A summary is provided in Table \ref{tab:n_rho}.  

\section{Results $\&$  Discussion}
\subsection{Brillouin Spectra and Mode Assignment}
\begin{figure}[t]
\includegraphics[scale=0.52]{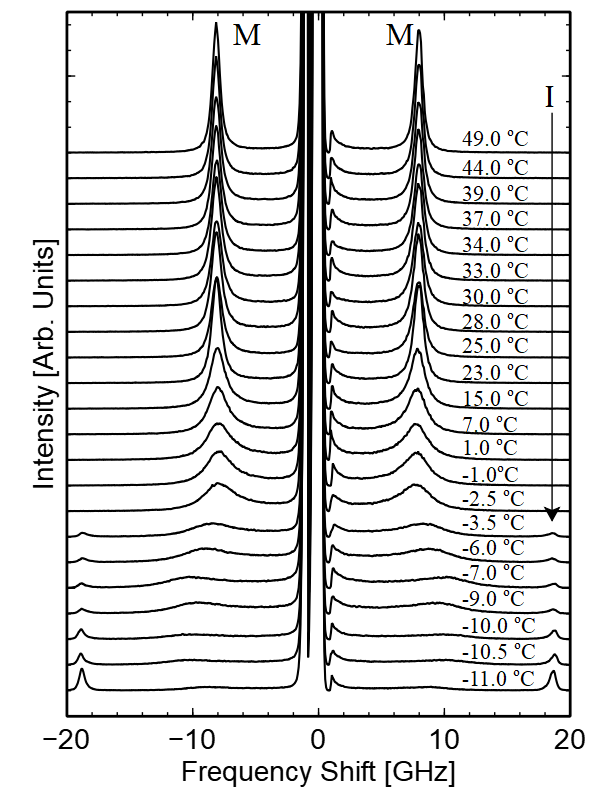}
  \caption{Brillouin spectra of a natural snail mucus (Sample \#2). Temperatures at which spectra were collected are indicated. \textbf{M} and \textbf{I} represent peaks due to mucus and ice, respectively.}
  \label{fig:BS_Sample2}
\end{figure}

Fig. \ref{fig:BS_Sample2} shows a representative series of mucus spectra obtained from one of three nominally identical samples at temperatures in the range -11$^\circ$C $\leq T \leq$ 52$^\circ$C.  Similar sets of spectra were collected from the other two samples (see Fig. S1 in the Supplementary Materials).  Peak parameters including frequency shifts, linewidths (FWHM), and integrated intensity were obtained by fitting Lorentzian functions to the Stokes and anti-Stokes peaks and averaging the resulting best-fit parameters.  To obtain intrinsic linewidths, the instrumental linewidth of 0.3 GHz was subtracted from FWHM values obtained from the Lorentzian fits. Estimated uncertainties in peak parameters were obtained from the uncertainty in the Lorentzian fits.  Raw peak parameter data (frequency shift, linewidth, and integrated intensity) for all three samples are presented in Tables S1-S3 in the Supplementary Materials.   

\begin{figure}[ht]
\includegraphics[scale=0.45]{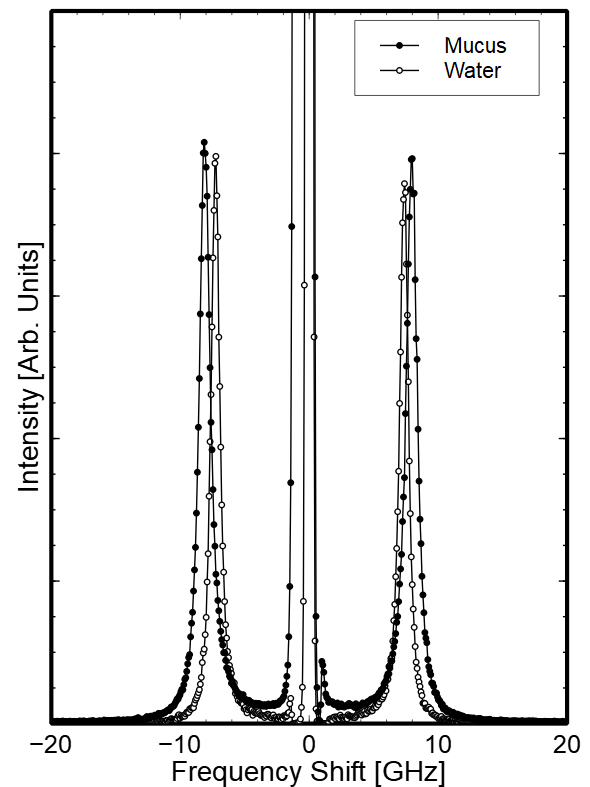}
  \caption{Room temperature Brillouin spectra of pure snail mucus and ultra pure water.}
  \label{fig:Mucus_Water_BS}
\end{figure}

Two sets of Brillouin peaks are present in the spectra - one at $\sim \pm8.0$ GHz over the entire temperature range and another at $\sim \pm18.5$ GHz for $T \leq$ -2.5$^\circ$C.  Due to the similarity of these shifts to those of liquid water (see Fig. \ref{fig:Mucus_Water_BS}) and that expected for solid water using a backscattering geometry at $T\lesssim0$ $^\circ$C \cite{gamm1980,gamm1983,gagn1990}, these peaks are attributed to the longitudinal acoustic mode in the liquid mucus and (likely polycrystalline) ice I$_h$, respectively. 

\subsection{Phase Transition: Raw Spectral Signatures}
\begin{figure}[!t]
\includegraphics[width=1\linewidth]{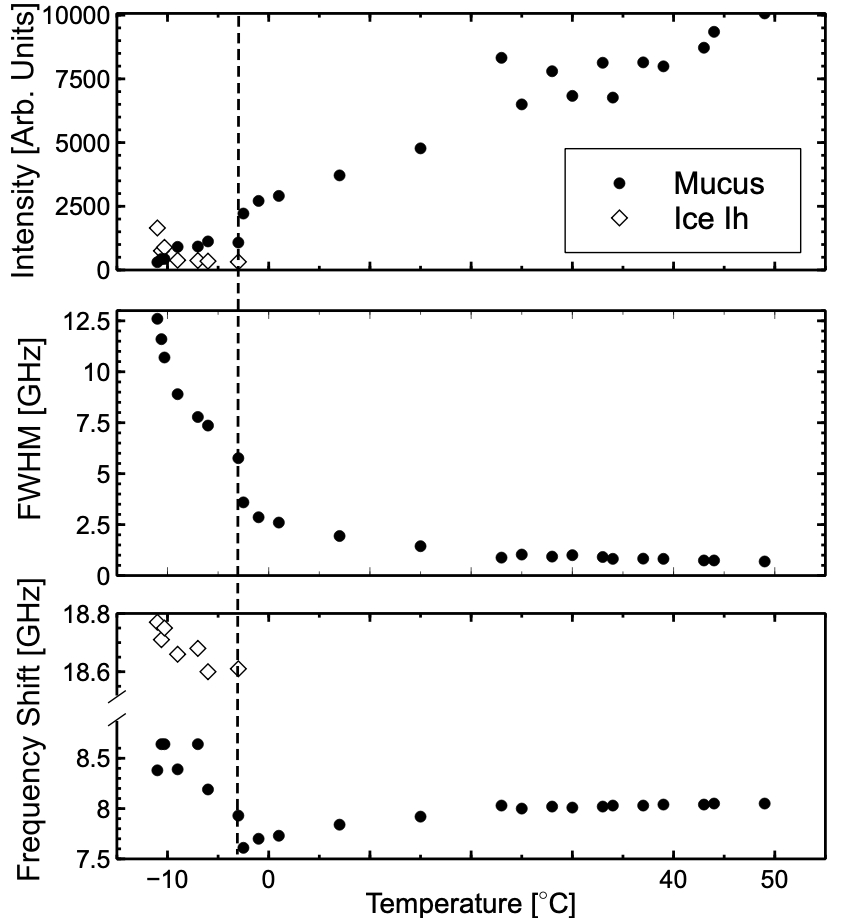}
  \caption{Brillouin peak intensity, linewidth, and frequency shift versus temperature for snail mucus. $\blacksquare$ - Mucus Mode, $\Box$ - Ice Mode.  The dashed vertical line indicates the phase transition at -2.5 $^{\circ}$C.}
  \label{fig:comp_plot}
\end{figure}

Fig.~\ref{fig:comp_plot} shows a discontinuous change in the temperature dependence of the mucus peak parameters at $T_{pt}=-2.5$ $^\circ$C and presence of the ice peak for $T\leq-2.5$ $^\circ$C, consistent with a transition of the system from liquid mucus to one in which mucus and ice coexist.  More specifically, the presence of a peak at a shift similar to that of the longitudinal acoustic mode for pure water in spectra collected for 50 $^\circ$C $\geq T \geq$ -2.5 $^\circ$C indicates that the mucus is in a liquid state over this range.  The frequency shift of this peak decreases gradually with decreasing temperature over this range. The intensity of this peak also shows an overall decrease with temperature and an abrupt drop near -2.5 $^\circ$C.  The peak width increases gradually with decreasing temperature over the same range, reflecting moderate damping of this mode as the temperature approaches -2.5 $^\circ$C.  Moreover, the absence of the ice peak for $T >-2.5$ $^\circ$C also indicates that only the liquid phase is present above this temperature. 

In the range -11.0 $^\circ$C $\leq T \leq$ -2.5 $^\circ$C, the presence of the peak due to liquid mucus and a second peak at a frequency shift close to that of polycrystalline ice Ih indicates the coexistence of liquid mucus with solid ice to at least -11.0 $^\circ$C.  The intensity of the ice peak increases with decreasing temperature over this range while there is a gradual decrease in the intensity of the liquid mucus peak, indicative of increasing amounts of ice and a reduction in the amount of liquid mucus in the system.  Over this range there is also a rapid increase in mucus peak linewidth with decreasing temperature, suggesting increased damping of this mode.  
Analogous sets of Brillouin data for two other nominally identical samples are shown in Fig. S2 in the Supplementary Materials document.
\begin{table*}
\caption{Hypersound velocity, storage (bulk) modulus, and loss modulus of a natural snail mucus and water at selected temperatures.  Quantities for the present work are average values determined from data for three nominally identical samples. Uncertainty in sound velocity is $\sim$ 2\%, 4\% in the bulk modulus and ranged from 3\% at 50$^\circ$C to 9\% at -11$^\circ$C.}
\begin{ruledtabular}
\begin{tabular}{ccccccccccccc}
Substance \& & \multicolumn{4}{c}{Velocity [m/s]} & \multicolumn{4}{c}{Storage (Bulk) Modulus [GPa]} & \multicolumn{4}{c}{Loss Modulus [GPa]}  \\
Temperature Range  & -10 $^\circ$C &  -2.5 $^\circ$C & 0 $^\circ$C & 50 $^\circ$C & -10 $^\circ$C &  -2.5 $^\circ$C & 0 $^\circ$C & 50 $^\circ$C & -10 $^\circ$C &  -2.5 $^\circ$C & 0 $^\circ$C & 50 $^\circ$C \\ \hline
Mucus, -11$^\circ$C $\rightarrow 50$ $^\circ$C [Pres Work] & 1770 & 1570 & 1580 & 1650 & 3.25 & 2.56 & 2.60 & 2.83 & 6.1 & 1.4 & 1.1 & 0.3 \\ 
Water, 0$^\circ$C $\rightarrow 50$ $^\circ$C [Pres Work] & -- & 1330 & 1360 & 1500 & - & 1.76 & 1.85 & 2.22 & - & 0.55 & 0.48 & 0.2\\
Water, -9 $^\circ$C $\rightarrow 100$ $^\circ$C - Ref. \cite{rouc1976} & 1280 & 1375 & 1395 & 1540 & 1.64 & 1.89 & 1.95 & 2.34 & - & - & - & - \\
\end{tabular}
\end{ruledtabular}
\label{tab:acoust_props}
\end{table*}
 
\subsection{Viscoelastic Properties}
Fig. \ref{fig:comp_plot2} shows the temperature dependence of the hypersound velocity, compressibility, and apparent viscosity derived from the Brillouin data for a representative sample (see Fig. S3 for corresponding data for the two other samples).  Viscoelastic properties at selected temperatures for this sample are presented in Table \ref{tab:acoust_props}. Also shown are viscoelastic properties for water determined in the present work along with those of water from previous studies. The behaviour of viscoelastic properties of these systems above and below the phase transition temperature of $T_{pt}=-2.5$ $^\circ$C is compared below. 
  
\subsubsection{Hypersound Velocity}
For 50 $^\circ$C $\geq T \geq$ -2.5 $^\circ$C, the mucus hypersound velocity shows an overall decrease at a rate of $\sim$ 1.5 m/s/$^\circ$C from $\sim1650$ m/s to $\sim1570$ m/s at the upper and lower ends of this interval, respectively (see Fig. \ref{fig:comp_plot}).  Analogous hypersound velocity data obtained for water in the present work and in previous studies \cite{teix1978,rouc1976,ocon1967} shows a similar trend over this temperature range, but with a rate of decrease that is $\sim2\times$ higher and magnitude that is $\sim10$\% lower than for the mucus.  Rates of decrease and velocities intermediate to those of water and snail mucus are observed for several aqueous polymer solutions with polymer concentrations similar to the glycoprotein concentration in the mucus over the temperature range 20 $^\circ$C to 45 $^\circ$C \cite{spic1988,spic1989,so1994,so1995,sirv1993,sirv1994,haqu1983}.   

For -2.5 $^\circ$C $\geq T \geq$ -11 $^\circ$C, the mucus hypersound velocity displays an overall increase at a decreasing rate from its minimum value of $\sim1570 $ m/s at $T_{pt}=-2.5$ $^\circ$C to $\sim1700$ m/s at -11.0 $^\circ$C.  This behaviour contrasts with that for supercooled water for which the velocity is $\sim10-30$\% lower and which exhibits a rapid monotonic decrease with temperature over this range \cite{teix1978,rouc1976}.  The formation of ice results in a larger glycoprotein-to-water ratio \cite{raym1977} resulting in higher mucus mode velocity which increases as temperature decreases due to additional ice formation. The ice mode velocity shows a slight increase from 3770 m/s at -2.5$^\circ$C to 3840 m/s at -11.0$^\circ$C.  The same trend is observed for pure polycrystalline ice Ih but the velocities are approximately 2-4\% lower than for the ice-like solid \cite{vaug2016}.  

\subsubsection{Complex Longitudinal Modulus \& Adiabatic Compressibility}
Values of mucus complex longitudinal modulus (storage and loss moduli) at select temperatures are presented in Table \ref{tab:acoust_props}. The storage (bulk) modulus exhibits a minimum at -2.5 $^\circ$C, a value that is approximately double that of pure supercooled water at the lowest temperature probed, and a value that is only $\sim20$\% greater than that for water at the highest temperature probed.  At $T_{pt} = -2.5$ $^\circ$C, the bulk modulus of mucus is $\sim35$\% larger than that for supercooled water. 
In general, for both mucus and ultra-pure water, the loss modulus decreases with increasing temperature. Furthermore, the value for both mucus and water at 50$^\circ$C are nearly identical. Like previous studies, the loss modulus in this system shows a greater change with temperature in its value than the storage modulus \cite{bail2020}. The likely reason for this is that water has a relatively high storage (bulk) modulus and the addition of a small amount of protein would not cause it to change significantly. The loss modulus,  shows analogous results to that of the FWHM as a function of temperature. Additionally, as the loss modulus is a measure of the amount of energy lost through heat in a system, we can say based on our data that as the temperature decreases more energy is lost in the system.

Fig. \ref{fig:comp_plot} shows the temperature dependence of the adiabatic compressibility of the mucus, $\kappa_s=1/\rho v^2$.  The compressibility displays a maximum at T = -2.5 $^\circ$C. Below $T_{pt}$, $\kappa_s$ decreases dramatically with decreasing temperature, while above the transition point it shows a gradual decline with increasing temperature. Interestingly, at the lowest temperature probed, the adiabatic compressibility of mucus is 50\% larger than that of water \cite{rouc1976} but is only $\sim$ 18\% higher than the value for water at the highest temperature probed.  Considering the bulk modulus values  reported for a previous study on water recorded in Table \ref{tab:acoust_props},  the data exhibit a general increase with increasing temperature for the entire temperature region probed. In contrast, our present work shows a general decrease in bulk modulus for $T \leq T_{pt}$, a minimum at the transition point, and then an overall increase for $T \geq T_{pt}$. The similarity in trends observed above the transition temperature can be attributed to the absence of ice. In this temperature range, the behaviour of the bulk modulus between mucus and water is the same. However, below the transition temperature, the discrepancy in bulk modulus values between mucus and water can be attributed to the presence of ice crystallites in mucus, which would significantly increase the rigidity of the system.

\begin{figure}[!t]
\includegraphics[width = 1\linewidth]{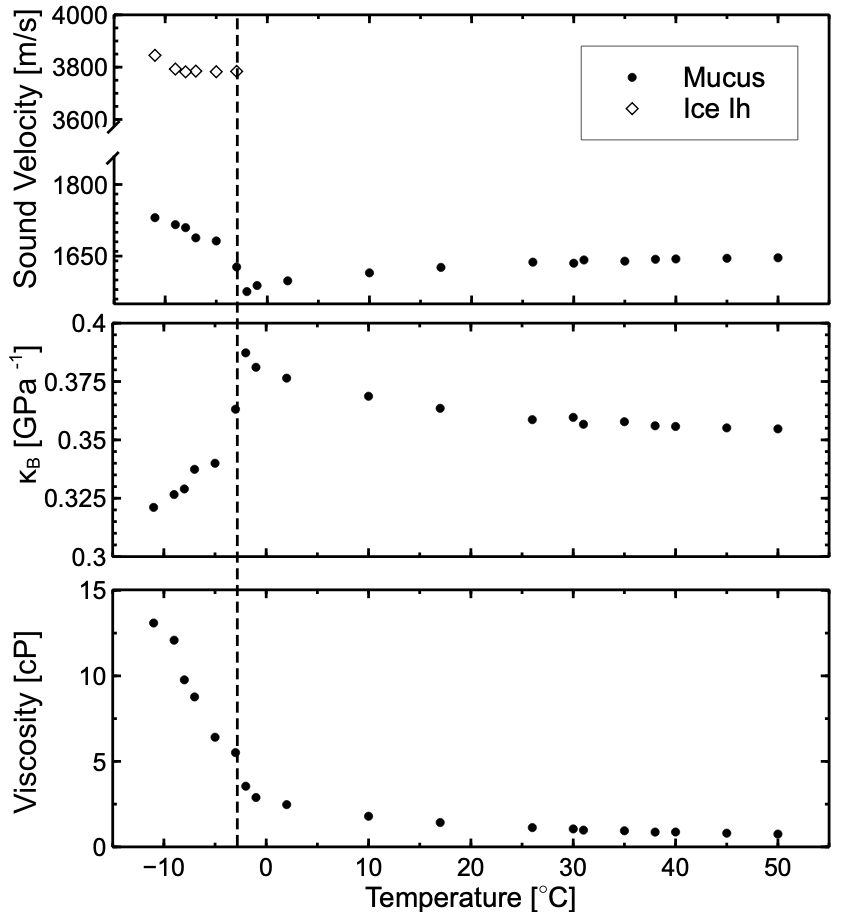}
  \caption{Hypersound velocity, adiabatic compressibility, and apparent viscosity versus temperature for a natural snail mucus (Sample \#2).  The dashed vertical line indicates the phase transition at -2.5 $^{\circ}$C.}
  \label{fig:comp_plot2}
\end{figure}

\begin{table}[!t]
\caption{Best-fit parameters for fit of function $\displaystyle \ln (\eta) = \ln \eta_0 +\frac{E_a}{k_B T}$ to experimentally determined apparent viscosity.}
\setlength{\extrarowheight}{3pt}
\begin{ruledtabular}
\begin{tabular}{c c c c c c}
Temperature & Sample & $\ln \eta_0$ & $E_a$ & R$^2$\\
Range &  &(cP) & kJ/mol &  \\
 \hline
&1& -4.5 $\pm$ 0.4 & 13.6 $\pm$ 0.3 & 0.987\\
25.0$^\circ$C $\leq T \leq$ 52.0$^\circ$C & 2 & -4.5$\pm$ 0.4 & 13.6 $\pm$ 0.3 & 0.993\\
&3&  -3.9 $\pm$ 0.6 & 13.5 $\pm$  0.5 & 0.983\\ \hline
& Mean & -4.3 $\pm$  0.5& 13.6 $\pm$ 0.4 & 0.988 \\  \hline
& Water\footnotemark[1] & -5.57 $\pm$ 0.3 & 13.8 $\pm$ 0.3& 0.991\\ 
    \end{tabular}
    \label{tab:ViscoFit}
    \end{ruledtabular}
    \footnotetext[1]{Ultra-pure water used in the current work.}
\end{table}

\subsubsection{Apparent Viscosity}
The temperature dependence of the apparent viscosity of the snail mucus obtained from Eqn. \ref{eq:viscosity} is shown in Fig.~\ref{fig:comp_plot2}.  The observed decrease with increasing temperature is similar to that previously reported for water \cite{holm2011} and other aqueous polymer solutions \cite{so1994,so1995,sirv1993,sirv1994,spic1989}. 
\begin{figure}[!t]
\includegraphics[scale=0.4]{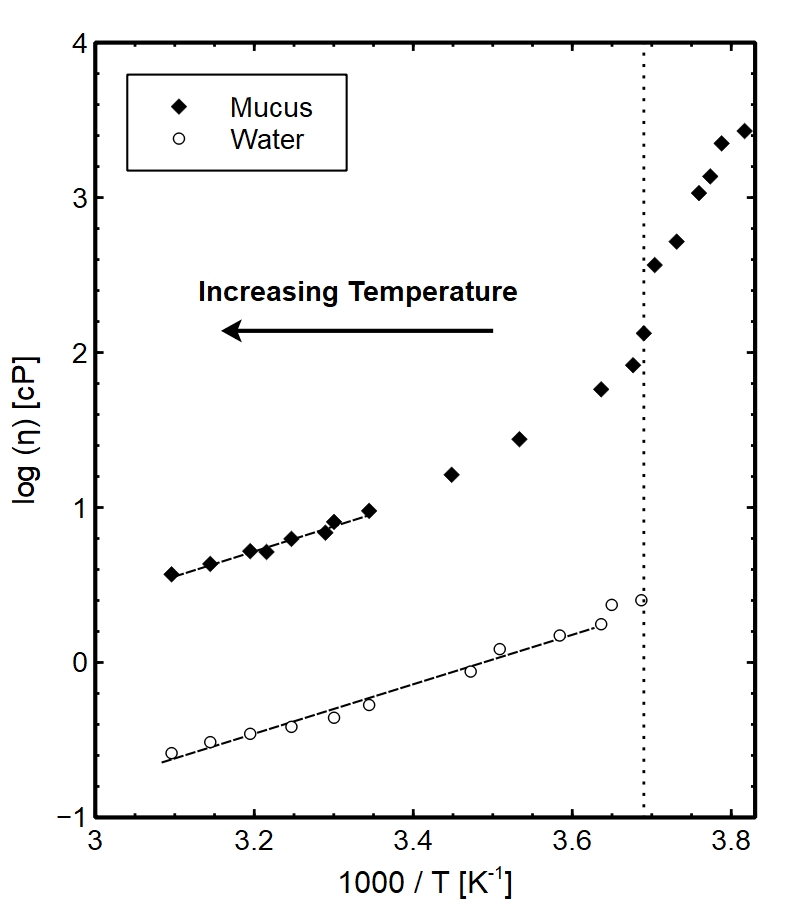}
  \caption{Natural logarithm of mucus viscosity versus inverse temperature. $\blackdiamond$ - Mucus, $\circ$ - Water. Dashed lines represents best fits. Results for linear fits are given in Table \ref{tab:ViscoFit}.  The dotted vertical line indicates the value of $1/T$ corresponding to the phase transition at -2.5 $^{\circ}$C.}
  \label{fig:logvisco}
\end{figure}

Fig. \ref{fig:logvisco} shows the natural logarithm of the apparent viscosity as a function of inverse temperature for snail mucus.  Analogous data for water is also shown for the purposes of comparison.  Over the high temperature range 25 $^\circ$C $\leq T \leq 52$ $^\circ$C, $\ln (\eta)$ depends linearly on $1/T$ so we fit an Arrhenius relationship of the form
\begin{equation}
\eta = \eta_0 e^{E_a/k_B T} 
\label{eq:lneta}
\end{equation}
to this data and extracted the activation (enthalpy) energy for the mucus $E_a^m = 13.6 \pm 0.4$ kJ/mol (H-bond values typically range from $\sim6-12$ kJ/mol \cite{brew2021,paol2007}).  Here, $\eta_0$ is a prefactor that contains the entropic contribution to the viscosity and $k_B$ is the Boltzmann constant \cite{lupi2011}.  A similar fit for water over the range 0 $^\circ$C $\leq T \leq 52$ $^\circ$C yields $E_a^w =13.8 \pm  0.3$ kJ/mol (see Table \ref{tab:ViscoFit}). 

The constant value of mucus activation energy over the high temperature range along with the fact that $E_a^m \simeq E_a^w$ over this range, suggests that the difference in apparent viscosity between mucus and water can be attributed to the entropic prefactor $\eta_0$. A higher value observed for mucus indicates an decreased level of disorder within the system, resulting in a lesser number of possible microscopic configurations compared to water. Moreover, due to viscosity being a measure of the internal friction of a fluid, the relatively higher apparent viscosity of mucus in this high-temperature region implies that mucus possesses a longer structural relaxation time in comparison to water. This observation is logical since an increase in the systems disorder or microscopic configurations would correspondingly require a longer duration for the system to relax. This is in line with previous studies \cite{benc2009,monaco2001glass,pochylski2005structural,bail2020,palo2019}, along with a proposed model consisting of a high-hydration state (liquid-like system) provides further support for this observation. Similarly, the apparent viscosity can also provide information on the density fluctuation relaxation times. The apparent viscosity, as mentioned earlier, represents the resistance to flow exhibited by a material. Since density fluctuations in a material are related to the internal molecular structure and motion of molecules in a system, it is therefore related to the viscosity inherently related to viscosity \cite{dil1982,comez2016,hansen2013theory}. In this high-temperature region, since the apparent viscosity is linear, this also suggests that the relaxation time for density fluctuation remains constant over temperature regime \cite{lupi2011}.

The intermediate region shows a nonlinear increase in $\eta$ up to the phase transition temperature $T_{pt}$. This increase in viscosity is accompanied by an increase in the structural relaxation time \cite{benc2009}. Moreover, it is important to note that bound and free water exhibit different density fluctuation relaxation times by default \cite{lupi2011}. Bound water, being more constrained in its motion near the solute, has slower dynamics compared to free water. As previously suggested these differences contribute to the change in apparent viscosity of the system, with their respective proportions of bound and free molecules playing a role \cite{lupi2011,meis2013}. Thus the non-Arrhenius behaviour seen in this region implies that the nonlinear increase in apparent viscosity is an artefact of density fluctuation relaxation time as well as structural relaxation times changing in a similar manner. Previous studies has stated that this non-Arrhenius behaviour is related to the onset of cooperative motions and molecular collisions of proteins and water molecules in the system \cite{comez2012progress}. In this temperature region, the non-Arrhenius behaviour is usually characterized by either the power law or the Vogel–Fulcher–Tammann equation \cite{comez2012,comez2012progress}. The latter is often associated with glass transitions in liquids.

Below the transition, there is a rapid, approximately linear, increase in the apparent viscosity. It is also worth noting that the values of $d\eta /dT$ for the mucus above and below $T_{pt}$ are drastically different. This is not surprising given that below $T_{pt}$ there is the coexistence of ice Ih and liquid mucus. A previous proposed model has suggested that in the case of antifreeze glycoproteins, that the glycoproteins are bound to water in the proteins respective hydration shell \cite{ebbi2010}. Furthermore, the data presented in this manuscript, combined with the apparent viscosity data, suggests a probable scenario where proteins bound to the ice require a higher activation energy, as observed here, to induce viscous flow. This can be attributed to the stronger bond existing between the protein, water, and ice.

\subsubsection{Hypersound Attenuation} 
\begin{figure}[!t]
\includegraphics[scale=0.35]{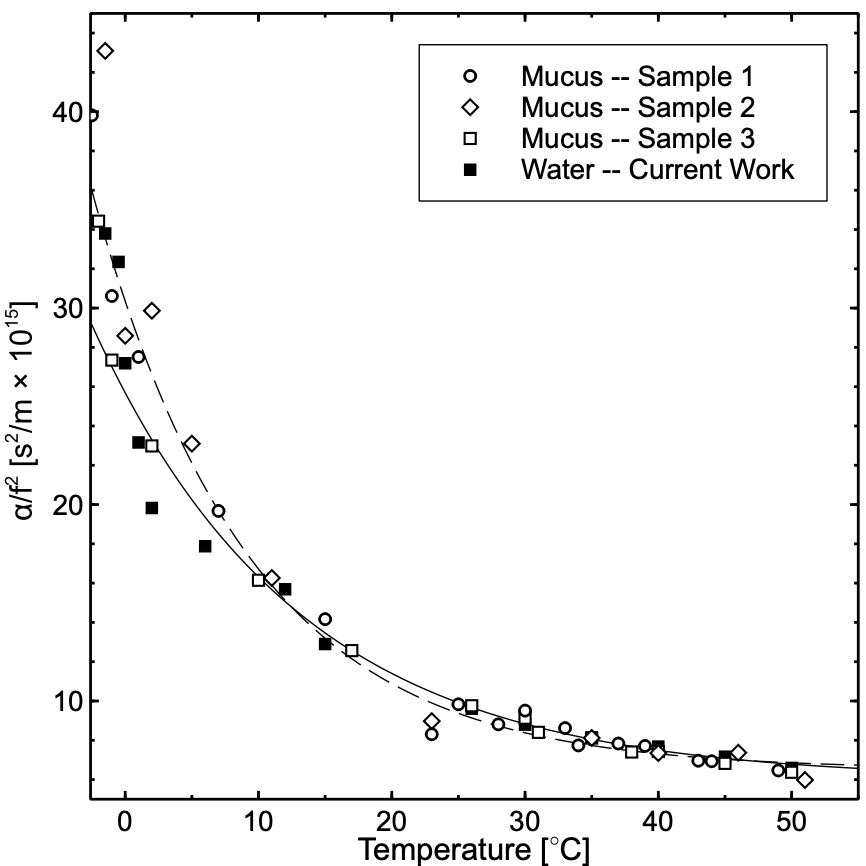}
  \caption{Plot of $\alpha/f^2$ for mucus samples and de-ionized water collected as a function of temperature over the range -2.5$^\circ$C $\leq T \leq$ 52.0$^\circ$C. Dashed line is a representative fit for all mucus samples. Solid line is the best fit obtained for ultra-pure water.}
  \label{fig:Soundabs}
\end{figure}

Fig. \ref{fig:Soundabs} shows the frequency-independent hypersound absorption coefficients for mucus and water determined from Eq. \ref{eq:absorp} as a function of temperature for $T > -2.5$ $^{\circ}$C. It is clear from this result that over this range more sound is absorbed in the medium as temperature decreases. This is expected because the absorption coefficient is proportional to apparent viscosity, which increases with decreasing temperature, and inversely proportional to hypersound velocity, which decreases with deceasing temperature (see Eq. \ref{eq:absorp}).  

The trend in Fig. \ref{fig:Soundabs} suggests fitting an equation of the form $\alpha/f^2 = A \exp(-B/T)$ to the \{$\alpha/f^2, T$\} data to obtain an understanding of how the sound waves in snail mucus are damped throughout the temperature range -2.5$^\circ$C $\leq T \leq$ 52.0$^\circ$C.  Table \ref{tab:EmpiricalFit} shows the best-fit parameters for both snail mucus and water in this work. The constant $A$ is a pre-exponential factor and a measure of the intensity of the absorption, while the parameter $B$ is a measure of how much sound is absorbed as a function of temperature. Additionally, $\epsilon$ derived from $B$ is a measure of the energy barrier associated with this sound damping.
 
\begin{figure}[!t]
\includegraphics[scale = 0.5]{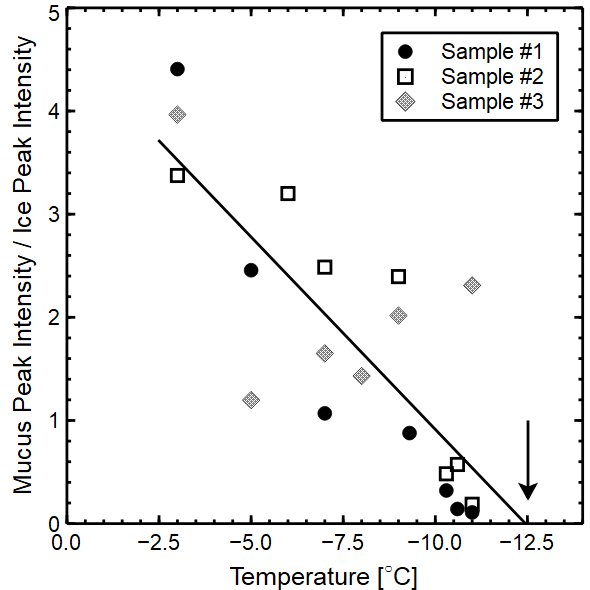}
\caption{Ratio of Mucus Peak Intensity to Ice Peak Intensity versus Temperature for three nominally identical samples of snail mucus. Solid Line - Line of best-fit to experimental data ($\mdlgblkcircle$, $\Box$, $\diamonddiamond$ ): $I_m$ / $I_i$ = 0.3731T + 4.6444. The arrow in the bottom right indicates the estimated temperature at which the mucus peak intensity equals zero.}
\label{fig:ImIi_T}
\end{figure}

\begin{figure}[!t]
\includegraphics[scale=0.4]{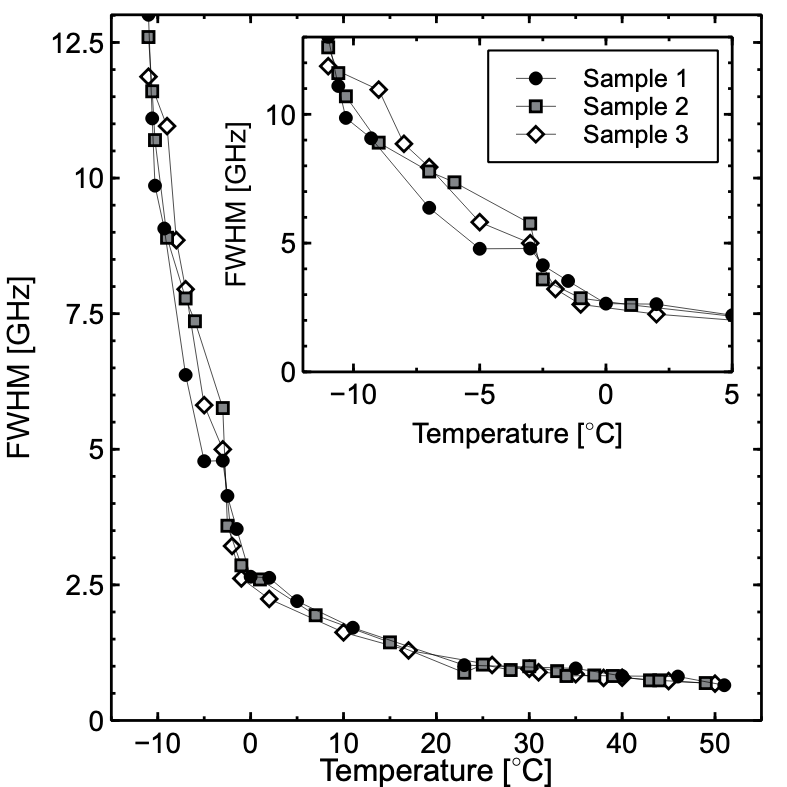}
\caption{Mucus Peak Linewdith versus Temperature.  The inset highlights the sample-to-sample differences in FWHM for temperatures below the phase transition at -2.5 $^\circ$C. }
\label{fig:FWHM_CompPlot}
\end{figure}

Below $T_{pt}$ one must also account for the possibility of additional attenuation effects due to the presence of ice crystallites.  The temperature dependence of $I_m/I_i$ shown in Fig. \ref{fig:ImIi_T} indicates that in most cases more water freezes into ice as temperature decreases, resulting in an increase in size and/or number of `obstacle' ice crystallites in solution.  There are, however, instances for which this is not the case, the behaviour varying from sample to sample, likely due to differences in the local microscopic environment, particularly with regard to glycoprotein concentration (see Fig. S2 in Supplementary Materials).  This, coupled with observed differences in mucus peak linewidth for the three samples for $T < -2.5$~$^\circ$C (see Fig. \ref{fig:FWHM_CompPlot}) suggests that damping of hypersound in this temperature range is not only due to absorption phenomena but also to scattering from ice crystallites in solution.

The non-zero intensity of the liquid mucus peak in spectra collected at -11 $^\circ$C (see Fig.~\ref{fig:BS_Sample2}) indicates that there is still coexistence of liquid and solid phases at the lowest temperature studied in this experiment. It is therefore not known if the mucus will solidify completely at a proximate lower temperature.  Linear extrapolation of the best-fit line for the $I_m/I_i$ versus temperature data shown in Fig. \ref{fig:ImIi_T}, however, suggests that complete freezing may occur at approximately -12.5$^\circ$C.  At this temperature, the liquid mucus mode is expected to be completely damped out, as suggested by the diverging peak FWHM at low temperature in Fig.~\ref{fig:FWHM_CompPlot}.

\begin{table}[!t]
\caption{Best-fit parameters for fit of function $\alpha/f^2 = A \exp(-B/T)$ to experimental sound absorption data for $T \geq -2.5$ $^\circ$C.}
\setlength{\extrarowheight}{3pt}
\begin{ruledtabular}
\begin{tabular}{c c c c c}
Sample &A & B& $\epsilon$  \\
&($\times 10^{-20}$ s$^2$/m) & ($\times 10^{3}$ K) & kJ/mol \\
 \hline
1& 1.1 $\pm$ 0.9  & 4.7 $\pm$ 0.5 & 39.6 $\pm$ 5.6\\
2& 3.1 $\pm$ 0.8  & 3.6 $\pm$ 0.6 & 29.9 $\pm$ 3.8 \\
3& 2.8 $\pm$ 0.7 & 3.3 $\pm$ 0.5 &  26.1 $\pm$ 3.8\\ \hline
Mean = & $2.3\pm 0.8$ & $3.9\pm 0.5$ & 31.9 $\pm$ 4.8 \\ \hline 
Water\footnotemark[1] & 0.7 $\pm$ 0.1 & 4.6 $\pm$ 0.6 & 38.6 $\pm$ 5.8 \\ 
Water \cite{rouc1976} & 0.5 $\pm$ 0.1 &3.7 $\pm$ 0.3 & 30.9 $\pm$ 5.8 \\ 
    \end{tabular}
    \label{tab:EmpiricalFit}
    \end{ruledtabular}
    \footnotetext[1]{Ultra-pure water used in the current work.}
\end{table}

\subsection{Phase Transition: Influence of Glycoproteins} \label{pt_gp}

\begin{figure}[!t]
\includegraphics[scale = 0.66]{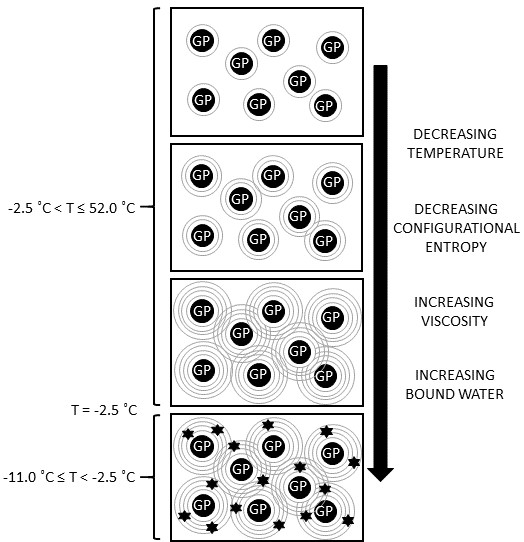}
\caption{Simplified schematic representation of changes in the snail mucus system with decreasing temperature over the range -11.0 $^\circ$C $< T \leq$ 52 $^\circ$C. $\CIRCLE$ - glycoprotein molecule, $\Circle$ - hydration shell, \SixStar\; - ice crystallite.  Due to reduced thermal energy, increasing amounts of water become bound to glycoprotein molecules (leaving less free water) and the hydration shells swell.  As a result, there is an increase in viscosity and a decrease in configurational entropy.  For -11.0 $^\circ$C $< T \leq$ -2.5 $^\circ$C, ice crystallites bound to glycoproteins are present in solution.}
\label{fig:schematic}
\end{figure}

The observed freezing point depression and coexistence of liquid and solid phases over a relatively large temperature range are manifestations of the inhibition of ice growth due to water-glycoprotein interaction.  On a microscopic scale, this effect is usually attributed to glycoprotein adsorption onto the surface of nucleating ice crystallites \cite{jia1996,knig1991}. Intuitively, however, a more likely track to freezing is one in which water binds to the mucus glycoprotein molecules at a temperature above the inception of freezing and, in fact, evidence for such binding has been observed for glycoproteins from {\it{Dissostichus Mawsoni}} \cite{ebbi2010}.  Bound water molecules in the glycoprotein hydration shell would have a reduced ability to reorient and take up a configuration that would permit freezing.  Due to a reduction in available thermal energy, increasing amounts of water would become bound to glycoprotein molecules as temperature is decreased, leaving less free water and causing hydration shells to swell.  This would result in a lowering of the temperature at which ice is formed and, when freezing did ensue under such conditions, the ice would be in the form of nanocrystallites attached to glycoprotein molecules in solution.  The results of the present study provide strong support for this route to freezing.  If the inhibition of ice growth was due to glycoprotein molecules adsorbing onto the surface of nucleating ice crystallites, one would expect to find the ice at the solution surface due to its density being lower than that of the liquid phase.  This is contrary to observations as we see definitive evidence for the presence of ice in solution in mucus Brillouin spectra in the form of a peak at a frequency shift corresponding to that of pure ice Ih. Furthermore, the non-negligible contribution of scattering to hypersound attenuation below the transition temperature is compelling indirect evidence that ice crystallites are present in solution.  This model of freezing is also supported by the results of THz absorption spectroscopy work which show that the diameter of the hydration shell of antifreeze glycoproteins from {\it{Dissostichus Mawsoni}} increases with decreasing temperature for 5 $^{\circ}$C $\geq T \geq 20$ $^{\circ}$C \cite{ebbi2010}. This, along with the fact that complexation of the glycoprotein with borate shifted the hydration dynamics towards that of free water, lead the authors to propose that long-range water-protein interactions contribute significantly to antifreeze activity.   Moreover, molecular dynamics simulations show that even an increase in hydration shell diameter of a few \AA\, can result in nearly all water being bound \cite{lupi2011,meis2013,cont2014}, thereby leaving little to no free water to freeze at or near the `equilibrium' freezing point.  It is conceivable, of course, that ice could be nucleating on the sides of the cell and that this is the source of our `ice' signal but this is unlikely given that its density is lower than that of liquid water and because we have three independent but nominally identical samples that behave essentially the same way.

\section{Conclusion}
The GHz-frequency viscoelastic properties and phase behaviour of a natural gastropod mucus were probed over the temperature range -11 $^\circ$C $\leq T \leq$ 52 $^\circ$C by Brillouin light scattering spectroscopy.   Anomalies in the temperature dependence of spectral peak parameters and derived viscoelastic properties reveal the inception of a previously uncharacterized liquid-to-solid phase transition at -2.5$^\circ$C, with coexistence of liquid mucus and solid ice persisting to at least -11 $^\circ$C.  The gradual nature of this transition and the observed temperature dependence of the apparent viscosity and hypersound attenuation is attributed to water-glycoprotein interaction, primarily that related to water binding to glycoprotein, reducing the ability of these bound water molecules to reconfigure and to crystallize to form ice.  This work gives new insight into role of bound water in biologically-important protein-water systems and also demonstrates the utility of Brillouin spectroscopy in the study of this unique class of natural materials, extending further the range of systems to which this technique has been productively applied. 

\section*{Acknowledgement}
\noindent
The authors would like to thank Gordon Whelan for constructing the temperature-controlled sample chamber.  GTA acknowledges the support of the Natural Sciences and Engineering Research Council of Canada (NSERC) (RGPIN-2015-04306).  

\bibliography{bibliography}

\end{document}